# A Sustainable Circular Framework for Financing Infrastructure Climate Adaptation: Integrated Carbon Markets

Chao Li[1], Xing Su*[1], Chao Fan[2], Jun Wang[3], Xiangyu Wang[4,5]

[1] College of Civil Engineering and Architecture, Zhejiang University, Hangzhou, Zhejiang, 310000, China;

[2] College of Engineering, Computing, and Applied Sciences, Clemson University, Clemson, SC, 29631, USA;

[3] School of Engineering, Design and Built Environment, Western Sydney University, Penrith NSW 2751, Australia;

[4] School of Civil Engineering and Architecture, East China Jiaotong University, Nanchang, Jiangxi, 330000, China;

[5] School of Design and the Built Environment, Curtin University, Perth, WA 6845, Australia.

**Abstract**

Climate physical risks pose an increasing threat to urban infrastructure, necessitating urgent climate adaptation measures to protect lives and assets. Implementing such measures, including the development of resilient infrastructure and retrofitting existing systems, demands substantial financial investment. Unfortunately, due to the unprofitability stemming from the long-term returns, uncertainty, and complexity of infrastructure adaptation projects and the short-term profit-seeking objectives of private capital, a massive financial gap remains. This study suggests incentivizing private capital to bridge financial gaps through integrated carbon markets. Specifically, the framework combines carbon taxes and carbon markets to involve infrastructure and individuals in the climate mitigation phase, using the funds collected for climate adaptation. Moreover, it integrates lifestyle reformation, environmental mitigation, and infrastructure adaptation to establish harmonized standards and provide circular positive feedback to sustain the markets. We further explore how integrated carbon markets can facilitate fund collection and discuss the challenges of incorporating them into infrastructure climate adaptation. This study aims to foster collaboration between private and public capital to enable a more scientific, rational, and actionable implementation of integrated carbon markets, thus supporting sustainable financial backing for infrastructure climate adaptation.

## 1. Introduction

Infrastructure affects 121 out of the 169 targets outlined in the 17 sustainable development goals (SDGs)[1]. In recent decades, the frequency and severity of climate physical risk factors, such as flooding and heat waves, have escalated[2,3], subjecting infrastructure to relentless and combined assaults[4,5]. These threats lead to the degradation and impairment of infrastructure, resulting in significant loss of life and property[6–8]. Therefore, urgent and crucial actions are needed to enhance the resilience of infrastructure to climate change and minimize losses in communities and cities [9,10].

Climate adaptation actions for infrastructure typically involve constructing resilient infrastructure and retrofitting existing systems[11], both types of adaptation necessitate substantial financial support. However, a significant gap exists between the demand for and supply of funds [12,13]. According to the Global Landscape of Climate Finance 2021-reported by the climate policy initiative, global climate adaptation funding only reached USD 46 billion in 2019/2020, which accounted for only 8% of climate mitigation funding, and the private sector contribution to climate adaptation funding was only 2%. The main reason for this disparity is the lower profitability of infrastructure adaptation compared to mitigation actions [14,15], as detailed in Section 2.1.

Existing literature on financing infrastructure adaptation primarily justifies the need and outlines top-level planning, such as the effectiveness and necessity of finance for climate adaptation[16], (inter)national systems and frameworks for climate adaptation finance[17], and financial instruments for climate adaptation [18]. However, it lacks in-depth research on the causes of the acute shortage of climate finance for adaptation and tailored financing solutions for infrastructure needs. Additionally, a systematic framework to ensure the sustained and appropriate use of climate finance for adaptation is notably missing.

To address these gaps, this study proposes a sustainable circular framework, integrated carbon markets, to bridge the financial shortfall. The integration of carbon markets is twofold. First, it entails the integration of lifestyle reformation, environmental mitigation, and infrastructure adaptation[19]. This integration is essential for providing sustained impetus to bolster infrastructure climate adaptation funds. A contrasting example is the reliance on complex bottom-up accounting for individual lifestyle changes to reduce carbon emissions, whereas governments typically employ a top-down approach to set carbon emission quotas, leading to inconsistent standards and confusing markets[20]. Second, it is critical to integrate mandatory (carbon taxes) and voluntary (carbon markets) measures in the carbon market, comprehensively addressing personal infrastructure-related carbon emissions. Carbon taxes establish clear emission reduction targets and norms for infrastructure operators, ensuring a minimum fiscal revenue for climate adaptation[21]. Carbon markets encourage a broader engagement in carbon reduction by motivating individuals and businesses towards low-carbon actions[22,23]. In integrated carbon markets, the accurate accounting of prices and allowance and the popularity of markets necessitate a deep understanding of the human-infrastructure-climate nexus. This study aims to identify new opportunities for understanding these interactions that arise from infrastructure climate adaptation, potentially enhancing private sector engagement through public awareness and economic incentives.

Fig.1a presents the vision of integrated carbon markets, featuring an external loop of lifestyle reformation, environmental mitigation, and infrastructure adaptation, as well as an internal core of carbon taxes and carbon markets. Infrastructures are categorized into seven types according to the present literature[1,24,25]. As energy, transportation, and waste contribute massive carbon emissions[26], these sectors not only participate in the carbon market but also need to pay additional carbon taxes (Fig.1b). This can serve two purposes. One is to accelerate the energy transition in the three sectors through punitive measures, and the other is to provide fundamental funds for integrated carbon markets when there is insufficient participation in the voluntary carbon market at the early stages. The rules for the use of funds are shown in Fig.1c. The carbon taxes, as the basic fund, is allocated to various critical infrastructure climate adaptations, while the funds collected by the carbon markets are mainly used for this type of infrastructure.

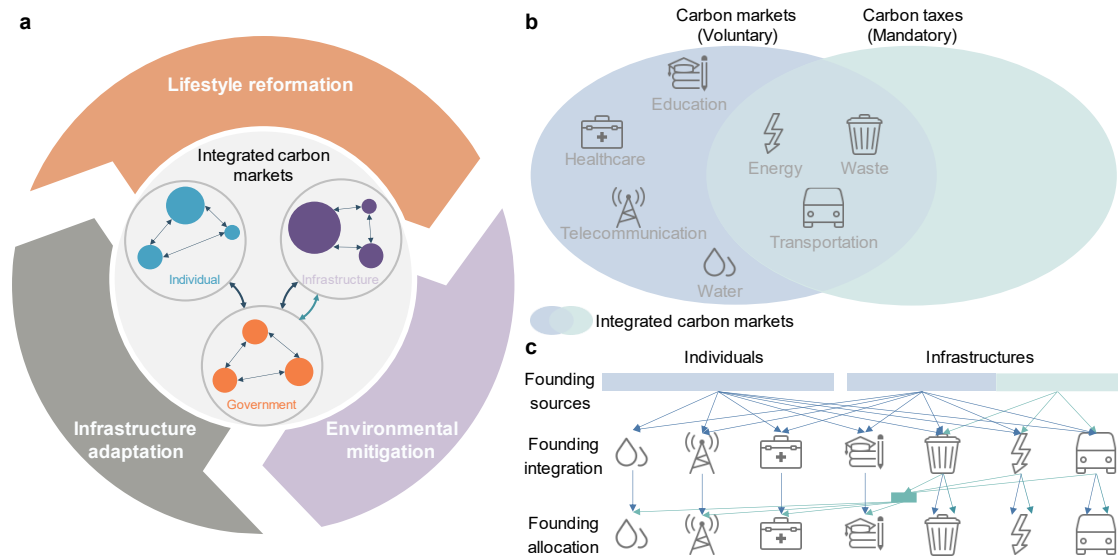

**Fig.1. a**) Conceptual framework of integrated carbon markets. It involves an external loop composed of lifestyle reformation, environmental mitigation and infrastructure adaptation, along with the internal core composed of integrated carbon markets. **b**) Categories of infrastructure and formation of integrated carbon markets. All categories of infrastructure are involved in the voluntary carbon market, and energy, waste and transportation infrastructure also pay a mandatory carbon tax because of their more massive carbon emissions. **c**) Routes for climate adaptation funding flows for a variety of infrastructure.

To the best of our knowledge, this study is the first to introduce the integrated carbon market to infrastructure climate adaptation by linking lifestyle reformation, environmental mitigation, and infrastructure adaptation into a cyclical process. Our framework creates a unified, systematic positive feedback loop that supports the sustainable operation of the integrated carbon market. Four fundamental research questions are addressed in this study: 1) Why is infrastructure climate adaptation underfunded? 2) How can integrated carbon markets bridge financial gaps? 3) What challenges arise when integrating carbon markets into infrastructure climate adaptation? 4) What are the foundational elements of this collaborative process?

The rest of the paper is organized as follows. In Section 2, This study thoroughly examined the factors that contribute to the lack of adequate climate finance for infrastructure adaptation, identifying unprofitability as a key barrier to attracting investment, particularly from the private sector. We then review existing climate financial instruments that are effective in attracting private capital, and find that the carbon pricing system has unique advantages (see Section 2.2 for details). In Section 3, based on the findings of Section 2, we propose a targeted framework for an integrated carbon market to bridge the financial gaps in infrastructure climate adaptation. We further demonstrate how integrated carbon markets can facilitate fund collection from three aspects. In Section 4, we discuss the challenges of integrating carbon markets to infrastructure climate adaptation. We offer several key foundations for implementing integrated carbon markets in Section 5 and concluding thoughts in Section 6.

## 2. Current status of infrastructure climate adaptation and climate financial instruments
## 2.1. Underfunding for infrastructure climate adaptation and its main causes

Driven by the continuing growth of cities around the world and the need to adapt to climate change, there is a tremendous need for climate finance on infrastructure. CCFLA provides a comprehensive assessment of the scale of funding required for these needs, with the cost of business-as-usual amounting to US$4.1-4.5 trillion per year[27]. In addition, the cost of adaptation adds another $120 billion (3%) per year, a figure that could be much higher if global average temperatures rise by more than 2°C compared to pre-industrial levels[28]. Even worse, governments seem to be more willing to allocate funds to climate mitigation than to climate adaptation. These actions further deepen the funding gaps for infrastructure climate adaptation.

Multiple factors have contributed to the historical neglect of infrastructure climate adaptation, with unprofitability serving as a prominent hindrance[14,15]. This unprofitability primarily stems from the significant uncertainty associated with climate physical risks[29]. The benefits of adaptation strategies are expressed in terms of "expected damages" avoided by the implementation of solutions[30], and that highly uncertain climate risks can lead to uncertain "expected damages". For instance, investing in low-carbon reusable materials during the construction phase of an infrastructure project can yield an immediate impact and generate revenue through market transactions by reducing carbon emissions[31]. In contrast, investing in resilience enhancements during the operational phase may only be validated after the onset of a climate hazard at an unexpected time, or may even be rendered insignificant due to the extreme intensity of the climate hazard.

The uncertain, opaque, and long-term benefits of climate change adaptation for infrastructure makes it an unappealing asset that fails to attract sufficient financial investment from both private and public capitals[32,33]. Numerous methods and techniques have been developed to underscore the advantages of climate change adaptation, including pricing climate risks for households and property owners[34–37]. Climate risk pricing entails integrating losses and damages induced by specific climate scenarios into an asset valuation model, facilitating the quantification of required investments and anticipated returns on climate adaptation financing[38]. Nonetheless, this method still covers only a limited number of disaster scenarios with significant uncertainties, making it difficult to attract investments needed to bridge the financial gap.

In addition, the inherent complexity of infrastructure systems can also deter profitability in climate adaptation efforts[39–41]. Infrastructure projects typically possess long asset lifespans and involve various stakeholders throughout their lifecycles, including governments, the private sector, and users[29]. The intricate responsibilities and demands placed upon each stakeholder lead to disparities in cost share and return allocation. A widely discussed example of such disparities is "free-rider" behavior, where individuals enjoy the collective benefits of climate adaptation equally without bearing their proportional costs[42]. Additionally, as cities expand, physical infrastructures grow in size and functionality over time, encompassing an increasing number of components and decades of diverse technologies[39]. These factors contribute to the mounting complexity of infrastructure systems. Therefore, the contribution of interdependent infrastructures to human well-being, the natural environment, and the economy in increasingly complex human systems is difficult to accurately quantify, and the massive indirect losses due to cascading effects in a disaster are difficult to define and account for[43,44]. Thus, calculating returns on investments in climate adaptation for infrastructure remains a challenge.

Due to these reasons, a persistent financial gap exists for infrastructure adaptation to climate change, especially in the short-term profit-driven private sector, despite some progress in addressing this issue.

**2.2. Existing global climate financial instruments**

The unprofitability of infrastructure climate adaptation projects has led to a failure to attract the interest of private capital. On the other hand, existing global policy practices have had more success in climate mitigation projects, with various financial instruments established to increase public and private sector involvement [45–51]. Among these, carbon pricing has become a key strategy for engaging the private sector[52]. Carbon pricing includes both indirect and direct approaches, with the former involving fossil fuel taxes and subsidies, while the latter comprises carbon taxes and carbon markets. Although indirect methods are more common, the adoption of direct carbon pricing mechanisms is growing [53]. As of April 2023, there are 73 operational carbon taxes or carbon markets worldwide, spanning regions such as the United States, Europe, and China[54]. Direct carbon pricing can be further categorized as mandatory or voluntary, depending on the willingness of participating entities. Mandatory initiatives include carbon tax (CT)[55] and emission trading system (ETS)[56]. These mechanisms aim to enforce stringent carbon emission management practices during the production of goods and services, with the funds collected being directed towards climate action. Voluntary approaches mainly involve personal carbon trading (PCT)[57], another type of carbon market mentioned earlier, designed to encourage low-carbon lifestyles among individuals, households, communities, and small and micro-enterprises through allowance trading and incentivized transactions within the carbon market [58].

However, existing mechanisms of carbon pricing focus primarily on reducing carbon emission for climate mitigation, overlooking climate adaptation[59]. They also pay insufficient attention to infrastructure sectors. Existing carbon tax and ETS frameworks predominantly target businesses involved in energy production and the manufacturing of industrial raw materials[60–62], such as steel and glass, paying less attention to direct carbon emissions from infrastructure construction and operation. Moreover, while PCT contributes to the reduction of consumption-based emissions by encouraging lifestyle changes[57], it fails to account for the mitigation of indirect carbon emissions resulting from infrastructure-related activities (i.e., personal emissions from infrastructure) prompted by individual behavioral changes.

In light of this, it is imperative to propose a tailored framework for financing infrastructure climate adaptation. Such a framework needs to build on existing advanced financial instruments for climate mitigation projects. Furthermore, integrating the characteristics of climate adaptation in different domains of infrastructure and their interactions with human activities is crucial for achieving targeted and actionable climate financing.

**3. How integrated carbon markets bridge financial gaps**

Capital, especially private capital, aims to maximize returns at minimum cost by utilizing available information[63,64]. This principle is challenged by investments in infrastructure climate adaptation, which are often neither informed nor profitable due to uncertainties about climate risks and the complexities of infrastructure systems. The integrated carbon market addresses these deep issues and the financial shortfall in infrastructure climate adaptation by providing adequate information and transparent incentives. The benefits of integrated carbon markets in bridging financial gaps manifest in three key ways.

**3.1. Offering transparent costs and returns for participants**

Fig.2b, Step 4 illustrates the conceptual process entailing carbon trading among individuals, infrastructure entities, and governments within the integrated carbon market. Notably, individuals and infrastructure operators assume dual roles as suppliers and demanders within the integrated carbon markets, while governments exclusively function as suppliers. To ensure transparent cost assessment, a clearly defined carbon tax rate is applied, outlining explicit costs related to energy, waste, and transportation infrastructure. Moreover, market-based carbon allowance trading operates by providing participants with transparent costs and returns through real-time carbon allowance prices that are determined based on the fundamental principles of supply and demand within the market.

Furthermore, a personal decision-making process for individuals in the market is introduced, as shown in Fig.2c. This delineates the diverse scenarios that individual participants may encounter within the carbon market, along with the corresponding actions they should undertake in response. It is worth noting that individuals are subject to carbon emission control, which encompasses both aggregate and unidirectional measures within the carbon market. Notably, when individuals face a shortage of carbon allowances, their primary recourse is to acquire allowances from other individuals, infrastructures, or the government. However, it is important to acknowledge that the transaction costs associated with these choices progressively increase. For instance, purchasing carbon allowances from the government incurs the highest costs.

The scenarios encountered by infrastructure participants within the carbon market and the corresponding response guidelines essentially mirror those depicted in Fig.2c, with the primary focus being on procuring carbon allowances from other infrastructure entities. Subsequently, allowances from individuals are considered, and ultimately, allowances from the government are sought as a last resort, albeit with the highest associated costs.

**3.2. Reinforcing knowledge about human-infrastructure-climate nexus**

The direct and indirect interactions between individuals and infrastructure play a crucial role in determining the resilience of infrastructure to climate change[65,66]. Understanding the complex dynamics of human-infrastructure-climate interactions is therefore essential for effectively allocating losses to infrastructure resulting from climate hazards and attributing the benefits derived from carbon emission reductions to individuals. This understanding serves to foster greater individual involvement in infrastructure climate action and narrow the financial gaps associated with infrastructure climate adaptation. Although significant progress has been made in comprehending the interrelationships between humans and infrastructure[67–70], there remains a dearth of research in quantifying the social, economic, and environmental impacts of infrastructure and allocating the costs and benefits of climate risk from the individuals' perspective.

Within our proposed framework, the initial step of integrated carbon markets (depicted in Fig.2b) entails a bottom-up allocation of carbon allowances. The successful implementation of this step hinges upon comprehensive research on human-infrastructure interactions. Such research should encompass essential characteristics of individuals, including their consumption patterns and travel behaviors, as well as the intrinsic attributes of infrastructure and the economic, social, and environmental implications it entails. The subsequent step integrates carbon allowances at the individual and infrastructure levels, categorizing carbon emissions from seven distinct types of individual consumption and accounting for direct and indirect emissions from infrastructures. The third step involves real-time carbon accounting, which enhances individual awareness of their carbon emissions. The combination of clear emission limits and heightened visibility serves to improve public perceptions of carbon emissions and, consequently, climate change. Additionally, a range of options exists for the climate adaptation of infrastructure, including enhanced public awareness and information disclosure, which can partially compensate for the lack of societal awareness regarding the interactions between people, infrastructure, and climate.

**3.3. Synergizing infrastructure climate actions**

Numerous initiatives have been undertaken to address climate change across the dimensions of lifestyle reformation[71–73], climate mitigation[74,75], and climate adaptation[76,77]. For instance, governmental efforts have focused on enhancing public awareness through education, thereby fostering lifestyle reformation[72]. Additionally, carbon quotas and credit systems have been implemented to reduce carbon emissions from enterprises, households, and individuals[78]. Climate adaptation has also been strengthened through the use of financial instruments like climate disaster insurance and climate funds, which help distribute and manage risks[79,80]. However, the inherent interconnections among these dimensions have often remained inadequately elucidated, and the absence of synergistic mechanisms has resulted in the isolation of these endeavors[81,82]. This isolation not only diminishes the efficiency and efficacy of climate action but also hinders the financing of infrastructure climate adaptation[83].

Our framework addresses this issue by first clarifying the interrelationships among these three dimensions and subsequently promoting their integration through the utilization of integrated carbon markets (depicted in Fig.2a). In this context, carbon markets employ financial incentives to encourage individuals to adopt low-carbon lifestyles. Such lifestyle reformation facilitates sustainable consumption patterns, thereby reducing indirect carbon emissions originating from infrastructure. Moreover, the combination of carbon taxes and carbon markets contributes to the reduction of direct carbon emissions from infrastructure. Consequently, reductions in carbon emissions from both individuals and infrastructure play a pivotal role in environmental mitigation. The outcomes of environmental mitigation serve as baseline scenarios for infrastructure adaptation, providing different carbon emission scenarios that guide diverse options for climate adaptation in infrastructure. Furthermore, governments generate funds through carbon taxes on priority sectors and the sale of carbon allowances in the carbon market. These funds can subsequently be allocated to support infrastructure climate adaptation actions. Various avenues exist for infrastructure climate adaptation, including enhancing physical resilience, public awareness, and information disclosure. These approaches help bridge gaps in societal perceptions and facilitate changes in individual lifestyles, thereby promoting effective climate adaptation.

By elucidating the interconnectedness of these endeavors and leveraging integrated carbon markets, our framework strives to enhance the integration of climate action across lifestyle reformation, climate mitigation, and climate adaptation. This integrated approach enables a more comprehensive and coordinated response to climate change, while also facilitating the financing of infrastructure climate adaptation efforts.

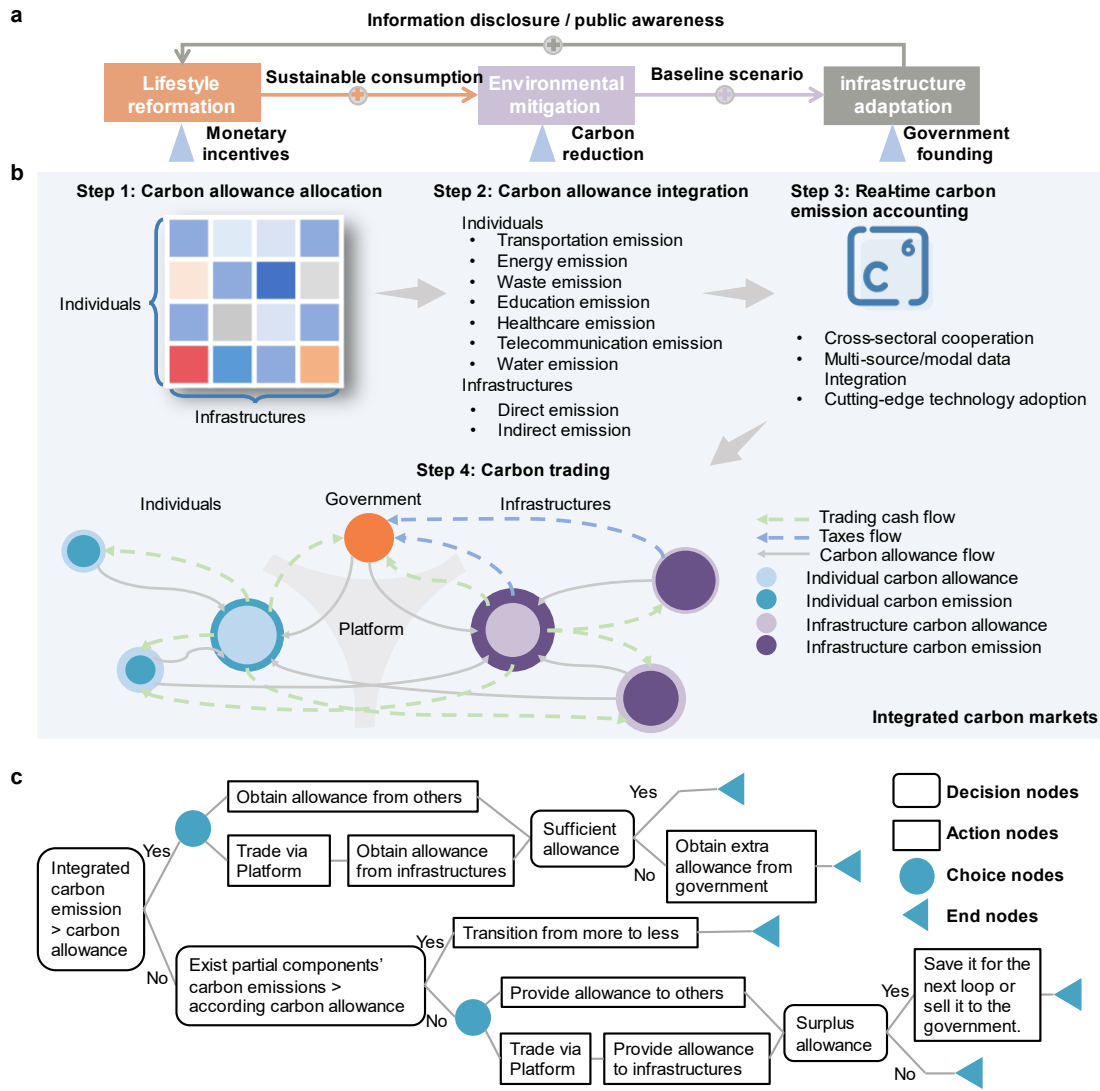

**Fig.2. a**) Contributions that the implementation of integrated carbon markets can make to synergize infrastructure climate actions. **b**) Core steps for integrated carbon markets and the conceptual process for carbon trading in integrated carbon markets. **c**) The personal decision tree when participates in integrated carbon markets.

## 4. Challenges of integrating carbon markets

The way to achieve a better collection of infrastructure adaptation funds is not by informing people to invest directly, but by establishing integrated carbon markets indirectly. This can help the public autonomously and actively explore relevant information and financial incentives, which in turn can help more successful public awareness. However, collaborating integrated carbon markets into infrastructure climate adaptation, while absolutely necessary, is not without challenges.

### 4.1. Inadequate support to provide transparent costs and returns

Accurate and real-time carbon accounting is an important basis for safeguarding transparent costs and returns in integrated carbon markets[57]. Hence, comprehensively capturing emissions from individuals' interactions with various infrastructures, as well as emissions from the activities of the infrastructures themselves are necessary conditions. This process requires exploring advanced methodologies and technologies, such as life-cycle assessments, consumer life methods and input-output analyses. Moreover, artificial intelligence and big data have great potential to effectively monitor and account for carbon emissions[84,85]. Through the use of sensors, the Internet of Things and extensive data analytics, it has become more feasible to monitor carbon emissions in real-time at a very fine scale, i.e., solo individual and infrastructure. However, more data and tools are not always better[86], so it is essential to clarify the data that needs to be collected, and the technologies and instruments that need to be utilized to account for real-time carbon emissions in advance.

In addition, an adequate funding pool provides a solid backing for offering transparent costs and returns to participants synchronously[87]. Getting more people to participate is the key to forming an adequate funding pool. Hence, it is imperative to discern the key drivers that underpin the long-term viability of the integrated carbon market. This inquiry can be pursued from two interconnected perspectives. Firstly, since meaningful participation in the carbon market hinges on inducing transformative changes in individuals' lifestyles, it is crucial to explore the fundamental drivers that motivate individuals to reform their way of life. Prior research in the field of personal carbon trading has identified three primary motivation factors[88]: social cognition, social norms[89], and economic incentives[90]. Notably, the relative contributions of these motivations exhibit substantial variation across individuals, necessitating an analysis tailored to individual characteristics. Secondly, the operational mechanisms governing the integrated carbon market can significantly influence its sustainable functioning. Several factors warrant careful consideration in this regard. For instance, the equitable allocation of carbon allowances, ensuring fairness among participants, assumes paramount importance. Moreover, the reliability and accuracy of carbon emissions measurement and reporting systems are vital to maintaining the integrity of the market. The affordability of carbon trading prices is another crucial aspect that must be taken into account to ensure broad and inclusive participation. Finally, the security and transparency of the carbon trading process are essential to foster trust and confidence among market participants.

**4.2. Insufficient theories on human-infrastructure-climate nexus**

The nexus encompassing human-infrastructure, human-climate, and infrastructure-climate dynamics has emerged as a persistent and pivotal area of research across past, present, and future studies[91]. The importance of advancing our understanding in this domain is equally critical within the context of integrated carbon markets, as highlighted earlier. The quantification of these interrelationships assumes particular significance when considering both pre-disaster and post-disaster scenarios[92].

In the pre-disaster phase, it is essential to quantitatively assess human-infrastructure interactions. This entails investigating the degree to which individuals' socioeconomic characteristics, travel behaviors, consumption patterns, and other factors are reliant upon corresponding infrastructure elements. Additionally, it is imperative to quantify the benefits and contributions of individual lifestyle reformation towards enhancing the resilience of infrastructure in the face of climate challenges. Furthermore, efforts should be directed towards quantifying infrastructure-climate interactions, including vulnerabilities and the significance of infrastructure in relation to climate risk factors characterized by various probabilities and severities. Understanding the potential social, economic, and environmental impacts resulting from diverse forms of physical and functional infrastructure damage is also essential in this context.

Following climate disasters, an accurate assessment of losses and damages is paramount. This involves evaluating the alignment between modelled outcomes projected under different pre-disaster scenarios and the actual post-disaster situation. Furthermore, it is crucial to quantify the extent to which adaptive investments in infrastructure mitigate losses and damages. Such quantification provides valuable insights for setting subsequent funding targets within the integrated carbon market framework.

By undertaking these quantitative investigations, we can deepen our knowledge of the intricate interplay between human activities, infrastructure systems, and climate dynamics. These insights will inform the development of robust strategies and policies for infrastructure climate adaptation, facilitating effective resource allocation within integrated carbon markets.

**4.3. Incomplete policy for infrastructure climate actions**

The successful implementation of synergizing infrastructure climate actions is contingent upon effective policy design and implementation[93]. Policymakers bear the responsibility of comprehensively examining the integrated carbon market and formulating a policy framework accompanied by relevant laws and regulations that align climate adaptation financing goals, markets, and policies. Furthermore, it is incumbent upon governments to establish a unified regulatory framework encompassing standardized accounting practices and rules for mutual recognition[94]. By assuming a guiding role, governments can effectively promote the healthy development of integrated carbon markets while ensuring market interoperability, fair competition, and the avoidance of policy overlap and conflicts. To achieve policy integration and coordination, the government should strengthen interdepartmental coordination through the establishment of an effective policy coordination mechanism[57].

The interests and concerns of each stakeholder within the integrated carbon market, encompassing environmental, monetary, and societal aspects, must be duly considered when calculating the equilibrium point under varying game scenarios. Additionally, the allocation of collected funds to different purposes can significantly impact the successful implementation of synergistic infrastructure climate actions within the integrated carbon market. Beyond the utilization of funds for the development of new resilient infrastructure and retrofitting existing infrastructure, considerations should extend to the potential use of funds for risk transfer through insurance purchases. Consequently, there is a pressing need to identify the optimal utility for fund allocation and strengthen the monitoring, reporting, and verifying (MRV) system.

By addressing these critical dimensions, policymakers can foster an environment conducive to the synergy of infrastructure climate actions. This entails aligning financial goals with comprehensive policy frameworks, establishing standardized regulations, promoting intergovernmental coordination, and ensuring the equitable consideration of stakeholder claims.

**5. Key foundations for implementing integrated carbon markets**

Advancements in science, data, and technology have the potential to expedite the integration of carbon markets with infrastructure climate adaptation. Within this evolutionary process, several key foundations play a particularly crucial role.

First, the co-creation, exchange and management of knowledge at all levels of government, private sectors, infrastructure managers and other relevant stakeholders need to be strengthened. There are multiple barriers to knowledge creation and dissemination, including knowledge about infrastructure climate adaptation and integrated carbon markets. The establishment of a unified information platform is essential to bridge the knowledge isolation of different stakeholders. Such a platform needs to be backed by comprehensive and unified information disclosure and universal data standards. Even more important is to allow the private sector and individuals to voice their opinions on well-considered interventions, propose changes, point out problems and suggest solutions.

Second, elements of the framework need to be tailored to local contexts. Given the wide-ranging variations in social, economic, and environmental conditions across countries and regions, the framework must exhibit the necessary flexibility to address diverse decision contexts, decision-makers, and actor groups. For instance, in low- and middle-income countries, it may be necessary to strengthen the proportion of carbon markets while reducing the reliance on carbon taxes. This adjustment aims to prevent the precipitous collapse of high-emission pillar industries.

Finally, potential negative consequences need to be avoided while attracting participations of private sectors. Equity and justice are more essential pillars for the public to engage in carbon markets compared to the effectiveness and performance of carbon markets[95]. Such as recognition justice (i.e., taking into account different sociocultural values), procedural justice (i.e., inclusiveness in decision-making), carbon allocation justice (i.e., the methodology for the allocation of initial carbon allowances) and funding allocation justice (i.e., the rational allocation of adaptation funds).

**6. Conclusion**

As infrastructure increasingly suffers from climate change impacts, it is facing substantial challenges in resilient actions. The lack of financial support is a common challenge at different levels of resilient actions. Although the United Nations, governments and sub-organizations have used a variety of financial instruments for climate finance, such as the green climate fund[96], the loss and damage fund[97] and catastrophe insurance[98], they are still inadequate in relation to the scale of funding required, especially for climate adaptation[99]. The uncertainty of climate physical risks and the complexity of infrastructure systems are the root causes of this problem. Considerable endeavors have been made to address these problems, yet achieving comprehensive solutions, especially in the short term, continues to be a daunting challenge. This situation underscores the urgent need to develop a new financing framework that effectively bridges the financial gap and mitigates the challenges associated with infrastructure climate adaptation.

We suggest a circular framework for infrastructure climate adaptation that revolves around integrated carbon markets. The integrated carbon markets can remediate problems by offering cost consideration and uncertain return for participants, enhancing knowledge about the human-infrastructure-climate nexus, synergizing infrastructure climate actions, and thus motivating the participation of the public and private sectors. The establishment of integrated carbon markets necessitates a deepened understanding of the complex dynamics within the human-infrastructure-climate nexus, an enhanced collection and integration of multi-modal and multi-source data, and the accelerated development and synergistic application of state-of-the-art digital technologies. It is crucial to acknowledge that the framework discussed here is still conceptual. The quantitative relationships between various framework components, such as specific proportions and values of carbon taxes and allowances allocated to particular infrastructures in specific regions, are not yet defined. Additionally, the extent to which these factors influence the willingness of infrastructure managers to engage in integrated carbon markets requires further investigation. Therefore, both novel research insights and an iterative application process are necessary to construct and refine the individual components within the proposed framework. Such endeavors will enhance the level of granularity of framework outputs, enabling a better comprehension of the complex interplay among different types of regions, infrastructures, population groups, and climate risk scenarios. Furthermore, these efforts will enable the framework to effectively respond to evolving infrastructure adaptation requirements and financial needs.

These frontier sciences and technologies, in combination with an actionable framework, will help to provide a broad range of decision-makers with the information to identify and remove barriers for disadvantaged populations, and to make choices and modifications that support equity and justice for participants.

44